\documentclass[english,english,pra,english,preprint,amsmath,amssymb,aps,longbibliography,showkeys,titlepage]{revtex4-2}
\usepackage{lmodern}
\usepackage{lmodern}
\usepackage[T1]{fontenc}
\usepackage[utf8]{inputenc}
\usepackage{color}
\usepackage{amsmath}
\usepackage{amsthm}
\usepackage{amssymb}
\usepackage{graphicx}
\usepackage{esint}

\makeatletter
\theoremstyle{plain}
\newtheorem{thm}{\protect\theoremname}
\theoremstyle{definition}
\newtheorem{defn}[thm]{\protect\definitionname}

\usepackage{amsthm}\usepackage{latexsym}\usepackage{bm}\usepackage{amsfonts}

\usepackage{babel}
\usepackage{hyperref}
\usepackage{enumitem}

\usepackage{hyperref}
\hypersetup{
	breaklinks = true,
    colorlinks = true,
    citecolor = {blue},
	urlcolor = {blue},
	linkcolor = {blue}
}

\@ifundefined{showcaptionsetup}{}{%
 \PassOptionsToPackage{caption=false}{subfig}}
\usepackage{subfig}
\makeatother

\usepackage{babel}
\providecommand{\definitionname}{Definition}
\providecommand{\theoremname}{Theorem}

\begin{document}
\title{Quantum Instability}
\author{Michael Q. May}
\email{mqmay@princeton.edu}

\affiliation{Plasma Physics Laboratory, Princeton University, Princeton, NJ 08540,
U.S.A}
\author{Hong Qin}
\email{hongqin@princeton.edu}

\affiliation{Plasma Physics Laboratory, Princeton University, Princeton, NJ 08540,
U.S.A}
\begin{abstract}
The physics of many closed, conservative systems can be described
by both classical and quantum theories. The dynamics according to
classical theory is symplectic and admits linear instabilities which
would initially seem at odds with a unitary quantum description. Using
the example of three-wave interactions, we describe how a time-independent,
finite-dimensional quantum system, which is Hermitian with all real
eigenvalues, can give rise to a linear instability corresponding to
that in the classical system. We show that the instability is realized
in the quantum theory as a cascade of the wave function in the space
of occupation number states, and an unstable quantum system has a
richer spectrum and a much longer recurrence time than a stable quantum
system. The conditions for quantum instability are described. 
\end{abstract}
\maketitle

\section{Introduction}

Many physical processes, in particular the dynamics of closed conservative
systems, may be usefully described by both quantum and classical theories.
However, classical dynamics are governed by Hamilton's equations whose
solutions are symplectic maps, while quantum dynamics, governed by
the Schrödinger equation, are unitary. A consequential difference
between symplectic maps and unitary maps is that the former allow
for dynamical instabilities and the latter do not. Traditionally,
this has required investigations into instabilities to occur solely
within the classical domain. For example, an opto-mechanical system,
where light in a resonant cavity is coupled to a mechanical arm which
affects the size of the cavity, can be described quantum mechanically
in terms of photons and phonons. But recent work dealing with the
opto-mechanical instability used a semi-classical approach rather
than the quantum description \citep{Ludwig_2008,bennett,rodriguez}.
This motivates the question of how to describe dynamical instabilities
using quantum mechanics. The issue of instabilities in quantum mechanics
has received some attention in the past, particularly concerning transitions
to chaotic behavior \citep{casati}, time-dependent Hamiltonians \citep{bellissard},
PT-symmetric Hamiltonians \citep{Bender1998,Bender2002,Bender2007,Mostafazadeh2002,Qin2019,Qin2021,Zhang2020},
and pseudo-Hermitian Hamiltonians \citep{ilya}. Despite these studies,
the fundamental question of how to characterize a quantum instability
in a time-independent, finite Hermitian Hamiltonian has been unaddressed.
In classical dynamical systems, a linear instability is characterized
by the existence of an eigenfrequency with positive imaginary part,
i.e, growth rate. This is impossible in a closed quantum system since
the dynamics are unitary, and its eigenfrequencies are real. However,
the unitary nature of Hermitian quantum systems does not preclude
the exponential growth of an observable that does not commute with
the Hamiltonian. In the present study, we define the concept of a
quantum instability as follows.
\begin{defn}
For a linear Hermitian quantum system with time-independent parameters,
an instability of an observable is a solution of the system for which
the expected value of the observable deviates exponentially from the
initial condition. When such an instability exits, the quantum system
is said to be unstable with respect to the observable. \label{def:1}
\end{defn}

In this paper, we will use the three-wave interaction model to to
illustrate how a such quantum instability occurs. Characterizing such
quantum instabilities is not only of academic interest, but also has
practical value for identifying when they can be realized or inevitably
occur on quantum systems. The three-wave interaction, which experiences
a classical instability, was recently simulated on a quantum computer
\citep{shiPhysRevA}. Although the simulation was not of high enough
dimension to allow for the instability, it demonstrates the importance
of the present work.

The three-wave interaction, the lowest order nonlinear interaction
in plasma dynamics, has applications in laser-plasma interactions
\citep{moody,myatt}, determining weak turbulence spectra \citep{zak2},
and nonlinear optical system design \citep{frantz,ahn,brunton}. Classically,
the linear dynamics of the interaction are affected by a parametric
instability before developing into the nonlinear regime \citep{Shi2018}.
Physically, this instability is triggered when one large amplitude
wave denoted by $(\omega_{1},k_{1})$ resonate with two others, denoted
by $(\omega_{2},k_{2})$ and $(\omega_{3},k_{3})$. In the so-called
decay interaction the resonance conditions are $\omega_{1}=\omega_{2}+\omega_{3}$
and $k_{1}=k_{2}+k_{3}$, which ensure energy-momentum conservation.
The instability transfers energy-momentum from the large wave to the
two smaller waves. Although this interaction and instability are well-known
\citep{rosenbluth,zakharov,Kaup1979,Reiman1979}, its quantum description
is less studied \citep{ohkuma,Shi2017,Shi2018,shiPhysRevA,Shi2021},
and the correspondence between the classical instability and its quantum
counterpart has not been established.

Section \ref{sec:The-Three-Wave} will focus on providing background
to the classical three-wave instability and the quantum theory of
three-wave interaction by Shi \textit{et al.} \citep{Shi2017,Shi2018,shiPhysRevA,Shi2021}.
In Sec.~\ref{sec:Quantum}, the quantum three-wave interaction equation
will be approximately solved in the linear regime, and an unstable
solution will be found, demonstrating the existence of a quantum instability
according to Definition 1. This unstable quantum solution will be
compared with the classical solution, and the classical limit of the
quantum instability will be discussed. Numerical results showing the
validity of the approximate linear solution of the quantum instability
will also be presented. In Sec.~\ref{sec:Numerical}, we will show
how the quantum instability is realized as a cascade of the wave function
in the space of the occupation number statues. The expected occupation
number of a quantum solution in terms of the eigenvalues of the Hamiltonian
will be derived. It shows a richer spectrum in the unstable system
than in the stable system. The eigenvalues of the stable Hamiltonian
are linearly distributed, while the eigenvalues of the unstable Hamiltonian
are nonlinearly distributed. Finally, Sec.~\ref{sec:Conclusions}
concludes with a discussion of the requirements for realizing the
quantum instability on quantum hardware and of future work to be done
on quantum instabilities.

\section{The three-wave interaction \label{sec:The-Three-Wave}}

In an ordinary gas, sound waves may nonlinearly self-steepen due to
interactions between the principle wave and its higher frequency resonances.
This is possible because each of the wave frequencies is a normal
mode of the system and therefore allowed. By contrast, most plasmas'
dispersion relations are very dispersive, so nonlinear interactions
of a single wave with its higher frequency resonances are negligible.
The lowest order nonlinear interaction in plasma dynamics is thus
the three-wave interaction, where a single wave interacts resonantly
with two others. For example, two Alfvén waves can interact nonlinearly
with a sound wave in a homogeneous plasma \citep{sagdeev}.

\subsection{Classical theory for three-wave interaction and instability}

In the classical theory, the nonlinear dynamics of the homogeneous
three-wave interaction may be reduced to \citep{jurkus,jaynes,Kaup1979,Reiman1979}
\begin{align}
\partial_{t}A_{1} & =gA_{2}A_{3},\label{eqn:a1}\\
\partial_{t}A_{2} & =-g^{*}A_{1}A_{3}^{*},\\
\partial_{t}A_{3} & =-g^{*}A_{1}A_{2}^{*},\label{eqn:a3}
\end{align}
where $g$ is a coupling coefficient, $A_{j}$ is the amplitude of
the $j$-th wave, and $A_{j}^{*}$ its complex conjugate. Equations
(\ref{eqn:a1})-(\ref{eqn:a3}) are the canonical Hamilton's equations
corresponding to the Hamiltonian

\begin{equation}
H=gA_{1}^{*}A_{2}A_{3}-g^{*}A_{1}A_{2}^{*}A_{3}^{*}\thinspace,\label{eq:H_classical}
\end{equation}
for the the canonical pairs of $A_{j}$ and $A_{j}^{*}$. By choosing
an appropriate normalization, we can let $g=1$ without losing generality.
The governing equations for the wave action $I_{j}=|A_{j}|^{2}$ are
found to be
\begin{align}
\partial_{t}I_{1}=-\partial_{t}I_{2}=-\partial_{t}I_{3}=gA_{1}^{*}A_{2}A_{3}+g^{*}A_{1}A_{2}^{*}A_{3}^{*}.\label{eqn:first_order}
\end{align}
These obviate two constants of motion in the system, 
\begin{align}
s_{2} & =I_{1}+I_{3},\label{eq:s2}\\
s_{3} & =I_{1}+I_{2},\label{eq:s3}
\end{align}
so that the growth of the second or third waves will reduce the amplitude
of the first. Using these constants of motion while taking another
time derivative of Eq.\,(\ref{eqn:first_order}), we arrive at closed
equations for the classical wave actions, 
\begin{align}
\partial_{t}^{2}I_{1}=\  & 2\left(s_{2}s_{3}+3I_{1}^{2}-2\left(s_{2}+s_{3}\right)I_{1}\right),\label{eqn:i1}\\
\partial_{t}^{2}I_{2}=\  & 2\left(s_{3}(s_{2}-s_{3})-3I_{2}^{2}+2\left(2s_{3}-s_{2}\right)I_{2}\right),\label{eqn:i2}\\
\partial_{t}^{2}I_{3}=\  & 2\left(s_{2}(s_{3}-s_{2})-3I_{3}^{2}+2\left(2s_{2}-s_{3}\right)I_{3}\right).\label{eqn:i3}
\end{align}
Note that Eq.\,(\ref{eqn:first_order}) implies that the right hand
sides of Eqs.\,(\ref{eqn:i1})-(\ref{eqn:i3}) are equivalent up
to a sign change. Equations (\ref{eqn:i1})-(\ref{eqn:i3}) are second-order
nonlinear differential equations which may be solved in terms of elliptic
integrals, and in the special case that $I_{2}=I_{3}$ the solutions
for $I_{1}$, $I_{2}$, and $I_{3}$ take on particularly simple forms
in terms of hyperbolic tangent and secant, respectively.

The linear instability of the three-wave system can be equivalently
described using Eqs.\,(\ref{eqn:a1})-(\ref{eqn:a3}) or Eqs.\,(\ref{eqn:i1})-(\ref{eqn:i3}).
For easy comparison with the quantum result in the next section, we
analyze the classical three-wave instability using Eqs.\,(\ref{eqn:i1})-(\ref{eqn:i3}).

In classical theory, linear instability refers to the exponential
growth of a deviation relative to an equilibrium solution of a system.
For the system studied here, the equilibrium solution of Eqs.\,(\ref{eqn:i1})-(\ref{eqn:i3})
is $I_{10}=\text{cont.}\neq0$ and $I_{20}=I_{30}=0.$ Consider a
perturbation of the system of the form,
\begin{align}
I_{1} & =I_{10}+\delta I_{1}\thinspace,\\
I_{2} & =\delta I_{2}\,,\\
I_{2} & =\delta I_{3}\,.
\end{align}
The linearized system for $\delta I_{1}$, $\delta I_{2}$, and $\delta I_{3}$
is
\begin{alignat}{1}
\partial_{t}^{2}\delta I_{1} & =0\thinspace,\label{eq:dI1}\\
\partial_{t}^{2}\delta I_{2} & =4I_{10}\delta I_{2}\thinspace,\\
\partial_{t}^{2}\delta I_{3} & =4I_{10}\delta I_{3}\thinspace.
\end{alignat}
Thus, the system is unstable with growth rate $\gamma=2\sqrt{I_{10}}.$
However, only $\delta I_{2}$ and $\delta I_{3}$ grow exponentially
with time. Equations (\ref{eqn:a1}) and (\ref{eq:dI1}) indicate
that for the unstable eigenmode of the linearized system, $\delta I_{1}$
and $\delta A_{1}$ remain constant. 

As will be shown in Sec.\,\ref{sec:Quantum}, an equilibrium solution
for the quantum system cannot be meaningfully defined. Therefore,
for comparison with the quantum solution, we derive the linear dynamics
of Eqs.\,(\ref{eqn:i1})-(\ref{eqn:i3}) relative to an initial condition
$I_{j}(0)=I_{ji}\neq0$ $(j=1,2,3)$. For exact solutions, $s_{2}$
and $s_{3}$ are conserved, so a solution for $I_{1}$ will determine
the solutions for $I_{2}$ and $I_{3}$. Assuming an initial condition
and a small deviation of the form
\begin{align}
I_{j} & =I_{ji}+\delta I_{j}\thinspace,(j=1,2,3)\,,
\end{align}
Eq.\,(\ref{eqn:i1}) can be rewritten as 

\begin{align}
\partial_{t}^{2}\delta I_{1}=2\left(\delta I_{1}(2(I_{1i}-I_{2i}-I_{3i})+3\delta I_{1})+I_{2i}I_{3i}-I_{1i}(I_{2i}+I_{3i})\right).
\end{align}
Assuming that $\delta I_{1}\ll\frac{2}{3}(I_{1i}-I_{2i}-I_{3i})$,
which will be true for short times, we may ignore the term proportional
to $\delta I_{1}^{2}$,
\begin{align}
\partial_{t}^{2}\delta I_{1}=2\left(\delta I_{1}2(I_{1i}-I_{2i}-I_{3i})+I_{2i}I_{3i}-I_{1i}(I_{2i}+I_{3i})\right).\label{eq:class_non_lin}
\end{align}
Its solution is 
\begin{equation}
\delta I_{1}=\frac{B_{C}}{\gamma_{C}^{2}}+C_{1}e^{\gamma_{C}t}-\left(\frac{B_{C}}{\gamma_{C}^{2}}+C_{1}\right)e^{-\gamma_{C}t}\,,
\end{equation}
where,
\begin{align}
\gamma_{C}= & 2\sqrt{I_{1i}-I_{2i}-I_{3i}}\,,\label{eqn:class_omg}\\
B_{C}= & 2I_{1i}(I_{2i}+I_{3i})-2I_{2i}I_{3i}\,.\label{eqn:class_gam}
\end{align}
The classical constants $\gamma_{C}$ and $B_{C}$ are written with
a subscript ``C'' to distinguish them from quantum constants $\gamma_{Q}$
and $B_{Q}$ which will be derived in the next section. The third
constant $C_{1}$ cannot be determined from the action equations,
Eq.\,(\ref{eq:class_non_lin}) and its counterparts for $I_{2}$
and $I_{3}$, because they each only involve even time derivatives
at the zeroth and second order. To determine $C_{1}$, the first order
Eqs.\,(\ref{eqn:a1})-(\ref{eqn:a3}) must be used. Choosing $(A_{1}(0),A_{2}(0),A_{3}(0))=(\sqrt{I_{1i}},\sqrt{I_{2i}},\sqrt{I_{3i}})$,
we find 
\begin{equation}
C_{1}=\frac{\sqrt{I_{1i}I_{2i}I_{3i}}}{\gamma_{C}}-\frac{B_{C}}{\gamma_{C}^{2}}.\label{eq:class_con}
\end{equation}
The growth rate $\gamma_{C}$ recovers the growth rate $\gamma$ derived
above when $I_{2i}=I_{3i}=I_{20}=I_{30}=0$. 

\subsection{Quantum theory for three-wave interaction}

The quantum theory for three-wave interaction is formulated by the
field-theoretical method, i.e. by quantizing the classical fields
$A_{j}$ as quantum operators $\hat{A}_{j}$ on the occupation number
states. The resulting Hamiltonian for a homogeneous (spatially independent)
quantum three-wave interaction with complex coupling constant $g$
is \citep{Shi2017,Shi2018,shiPhysRevA,Shi2021}
\begin{align}
\hat{H}=ig\hat{A}_{1}^{\dagger}\hat{A}_{2}\hat{A}_{3}-ig^{*}\hat{A}_{1}\hat{A}_{2}^{\dagger}\hat{A}_{3}^{\dagger},\label{eqn:ham}
\end{align}
where the $\hat{A}_{j}^{\dagger}$, $\hat{A}_{j}$ $(j=1,2,3)$ are
creation and annihilation operators, respectively, and $[\hat{A}_{j},\hat{A}_{l}^{\dagger}]=\delta_{jl}$.
This Hamiltonian acts on the space of occupation number states $|n_{1},n_{2},n_{3}\rangle$,
and commutes with operators $\hat{s}_{2}=\hat{n}_{1}+\hat{n}_{3}$
and $\hat{s}_{3}=\hat{n}_{1}+\hat{n}_{2}$, where the $\hat{n}_{j}=\hat{A}_{j}^{\dagger}\hat{A}_{j}$
are standard number operators. The $\hat{s}_{2}$ and $\hat{s}_{3}$
operators are identical in form to the constants of motion found in
the classical theory, Eqs.\,(\ref{eq:s2}) and (\ref{eq:s3}), so
we will write their expectation values identically, i.e. $\langle\hat{s}_{2}\rangle=s_{2}$
and $\langle\hat{s}_{3}\rangle=s_{3}$.

Because $\hat{s}_{2}$ and $\hat{s}_{3}$ commute with $\hat{H}$,
their eigenstates form an invariant subspace of dimension $d=s_{2}+1$
\citep{shiPhysRevA,Shi2021} with states

\begin{equation}
\Psi(t)=\sum_{i=0}^{s_{2}}\alpha_{i}(t)\psi_{i},\label{eqn:psi}
\end{equation}
where,
\begin{equation}
\psi_{i}=\left|s_{2}-i,s_{3}-s_{2}+i,i\right\rangle .\label{eqn:psi2}
\end{equation}
It is assumed that the eigenvalue $s_{3}\geq s_{2}$, which accounts
for the asymmetry in the above equation. Within this subspace, the
Hamiltonian is represented as a square tridiagonal matrix with vanishing
diagonal. Taking the coupling constant $g=-i$, the matrix is also
symmetric, with elements
\begin{align}
H_{ij}=\delta_{i,j+1}h_{i}+\delta_{i+1,j}h_{j}\thinspace,\label{eq:Hij}\\
h_{i}=\sqrt{(s_{2}-i)(s_{3}-s_{2}+1+i)(i+1)}.\label{eqn:h}
\end{align}
As will be shown below, the phase of the coupling constant will not
affect the dynamics of observables. Also, we emphasize that this quantization
procedure using the field-theoretical method maps the classical nonlinear
Hamiltonian specified by Eq.\,(\ref{eq:H_classical}) into a quantum
(linear) Hamiltonian operator on a finite-dimensional Hilbert space. 

Recently, the quantum three-wave interaction was simulated by Shi
\emph{et al}. \citep{shiPhysRevA} on a Rigetti Computing hardware
using a system with $d=3$. By realizing the unitary operator as a
single gate, they were able to robustly simulate the quantum three-wave
interaction an order of magnitude longer than by approximating the
unitary operator as a series of native gates.

Returning to Eq.\,(\ref{eqn:ham}) with an arbitrary coupling coefficient
$g$, the Heisenberg equations are
\begin{align}
\partial_{t}\hat{A}_{1} & =g\hat{A}_{2}\hat{A}_{3},\nonumber \\
\partial_{t}\hat{A}_{2} & =-g^{*}\hat{A}_{1}\hat{A}_{3}^{\dagger},\nonumber \\
\partial_{t}\hat{A}_{3} & =-g^{*}\hat{A}_{1}\hat{A}_{2}^{\dagger},\label{eqn:heisen}
\end{align}
which are identical in form to the amplitude equations of the classical
case, Eqs.\,\textcolor{black}{(\ref{eqn:a1}})-(\textcolor{black}{\ref{eqn:a3}}),
with classical wave amplitudes replaced by operators. Also as with
the classical case, we may combine these equations using the constants
of motion to find decoupled second-order equations for the number
operators $\hat{n}_{j}=\hat{A}_{j}^{\dagger}\hat{A}_{j}$,
\begin{align}
\partial_{\tau}^{2}\hat{n}_{1}=\  & 2\left(\hat{s}_{2}\hat{s}_{3}+3\hat{n}_{1}^{2}-\left(2\hat{s}_{2}+2\hat{s}_{3}+1\right)\hat{n}_{1}\right),\label{eqn:n1}\\
\partial_{\tau}^{2}\hat{n}_{2}=\  & 2\left(\hat{s}_{3}(1+\hat{s}_{2}-\hat{s}_{3})-3\hat{n}_{2}^{2}+\left(4\hat{s}_{3}-2\hat{s}_{2}-1\right)\hat{n}_{2}\right),\label{eqn:n2}\\
\partial_{\tau}^{2}\hat{n}_{3}=\  & 2\left(\hat{s}_{2}(1+\hat{s}_{3}-\hat{s}_{2})-3\hat{n}_{3}^{2}+\left(4\hat{s}_{2}-2\hat{s}_{3}-1\right)\hat{n}_{3}\right),\label{eq:n3}
\end{align}
where $\tau=t|g|$, and $\partial_{\tau}^{2}\hat{n}_{1}=-\partial_{\tau}^{2}\hat{n}_{2}=-\partial_{\tau}^{2}\hat{n}_{3}$.
Note that Eqs.\,(\ref{eqn:n1})-(\ref{eq:n3}) depend only on the
magnitude of the coupling constant, which has been absorbed by the
normalized time parameter. Next, to fairly compare the quantum and
classical equations, we take the expectation of Eqs.\,(\ref{eqn:n1})-(\ref{eq:n3}),
\begin{align}
\partial_{\tau}^{2}\langle n_{1}\rangle=\  & 2\left(s_{2}s_{3}+3\langle n_{1}^{2}\rangle-\left(2s_{2}+2s_{3}+1\right)\langle n_{1}\rangle\right),\label{eqn:quant_main}\\
\partial_{\tau}^{2}\langle n_{2}\rangle=\  & 2\left(s_{3}(1+s_{2}-s_{3})-3\langle n_{2}^{2}\rangle+\left(4s_{3}-2s_{2}-1\right)\langle n_{2}\rangle\right),\label{eq:quant_main2}\\
\partial_{\tau}^{2}\langle n_{3}\rangle=\  & 2\left(s_{2}(1+s_{3}-s_{2})-3\langle n_{3}^{2}\rangle+\left(4s_{2}-2s_{3}-1\right)\langle n_{3}\rangle\right).\label{eq:quant_main3}
\end{align}
Directly comparing the classical Eqs.\,(\ref{eqn:i1})-(\ref{eqn:i3})
with their quantum counterparts defined above, we see that as with
the classical case, the right hand sides of Eqs.\,(\ref{eqn:n1}),
(\ref{eqn:n2}), and (\ref{eq:n3}) differ only by their signs. There
are also significant differences. Each equation now includes additional
factors of the number of photons $\langle n_{j}\rangle$ on the right
hand side of the equations, and the last two equations also include
additional constant factors. More importantly, the quantum and classical
second-order equations differ in that the quantum equations are not
closed equations for photon number $\langle n_{j}\rangle$. They also
depend on the variance $\delta_{j}=\langle n_{j}^{2}\rangle-\langle n_{j}\rangle^{2}$.
As will be discussed in Sec.\,\ref{sec:Quantum}, the variance cannot
be zero for all times except in the trivial solution $\langle n_{1}\rangle=\langle n_{2}\rangle=\langle n_{3}\rangle=0$.
Since the variance is nonzero, either a closure must be established
for Eqs.\,(\ref{eqn:quant_main})-(\ref{eq:quant_main3}) to be useful,
or the full Schrödinger equation must be solved numerically. 

\section{The Quantum theory of three-wave Instability\label{sec:Quantum}}

In this section, we find an approximate solution for $\langle n_{1}\rangle(\tau)$
which corresponds to the exponential growth of the unstable classical
solution as discussed in Sec.\,\ref{sec:The-Three-Wave}. This first
requires assuming an initial condition so as to determine the variance
$\delta$$_{j}$. With an expansion of the variance, we then linearize
Eq.\,(\ref{eqn:quant_main}) and find the growth rate of the quantum
instability. The unstable solutions according to the classical and
quantum descriptions of the three-wave interaction are compared. Finally,
the approximate solution is validated by the numerical solution of
the Schrödinger equation. 

\subsection{Approximate solution of the variance}

We seek to solve Eq.\,(\ref{eqn:quant_main}), which depends on the
variance $\delta_{1}$. Estimating $\delta$$_{1}$ requires considering
the Schrödinger equation
\begin{equation}
i\partial_{\tau}\Psi=H\Psi,\label{eq:schr}
\end{equation}
where the Hamiltonian $H$ is defined in Eq.\,(\ref{eqn:ham}) and
the constant $\hbar=1$. The Hamiltonian for the invariant subspace
of constant $s_{2}$ and $s_{3}$ is a $d\times d$ matrix,
\begin{eqnarray}
H=\left(\begin{array}{cccccc}
0 & h_{0} & 0 & 0 & 0 & \dots\\
h_{0} & 0 & h_{1} & 0 & 0 & \dots\\
0 & h_{1} & 0 & h_{2} & 0 & \dots\\
0 & 0 & h_{2} & 0 & h_{3} & \dots\\
\vdots & \vdots & \vdots & \vdots & \vdots & \ddots
\end{array}\right),
\end{eqnarray}
where $h_{j}$ are defined by Eq.\,(\ref{eqn:ham}). Equation (\ref{eq:schr})
is a system of $d$ coupled first-order differential equations, 
\begin{align}
i\dot{\alpha}_{0} & =h_{0}\alpha_{1},\label{eqn:w1-1}\\
i\dot{\alpha}_{1} & =h_{0}\alpha_{0}+h_{1}\alpha_{2},\\
 & \ \ \ \ \ \dots\nonumber \\
i\dot{\alpha}_{i} & =h_{i-1}\alpha_{i-1}+h_{i}\alpha_{i+1}.\label{eqn:w2-1}
\end{align}
which are written explicitly in terms of the basis vectors $\psi_{i}$
defined in Eqs.\,(\ref{eqn:psi}) and (\ref{eqn:psi2}). As basis
vectors in this $d$ dimensional space, $\mathbf{\psi}_{i}$ can be
represented as 
\begin{equation}
\mathbf{\psi}_{i}=(\underbrace{0,0,\dots0,}_{i}1,0,0,\dots0).\label{eq:basis}
\end{equation}

As a footnote, we point out that it is straightforward to show by
construction that there exists a unique nontrivial equilibrium solution
such that $\dot{\alpha_{i}}=0$ for all $i$. However, this zero-energy
eigenstate should not viewed as the quantum counterpart of the equilibrium
in the classical theory.

The variance of the observable $\hat{n}_{1}$ is 

\begin{align}
\delta & =\langle n_{1}^{2}\rangle-\langle n_{1}\rangle^{2}\nonumber \\
 & =\sum_{j=0}^{s_{2}}|\alpha_{j}|^{2}\left(s_{2}-j\right)^{2}-\left(\sum_{j=0}^{s_{2}}|\alpha_{j}|^{2}(s_{2}-j)\right)^{2}.\label{eqn:del_def}
\end{align}
Note that for an infinitely narrow initial condition, where $\alpha_{m}(0)=1$
and $\alpha_{i\neq m}(0)=0$ for some $m$, the variance is zero.
Using the above expression of $\delta_{1}$, it can also be proven
that its maximum value is $s_{2}^{2}/4$. 

For a sufficiently narrow distribution of initial states, it may be
possible to approximate the variance as a constant, so long as it
does not grow too quickly in time. To justify this approximation,
consider a narrowly distributed initial condition 
\begin{align}
\Psi(0)=(\alpha_{0}(0),\alpha_{1}(0),\dots,\alpha_{s_{2}}(0))=(\dots,\phi\varepsilon^{2},\phi\varepsilon,\phi,\phi\varepsilon,\phi\varepsilon^{2},\dots),\label{eq:init_spread}
\end{align}
where $\phi$ is a normalization such that $\sum_{i=0}^{s_{2}}|\alpha_{i}(0)|^{2}=1$,
and $\varepsilon\ll1$ is a small parameter describing how spread
out the initial state is. Algebraically, this requires that for some
$m$, $\alpha_{i}(0)=\varepsilon^{|m-i|}\alpha_{m}(0)$, i.e., the
initial distribution is centered around the $m$-th state $|n_{1},n_{2},n_{3}\rangle=|s_{2}-m,s_{3}-s_{2}+m,m\rangle$.
At $\tau=0$, the variance of $\hat{n}_{1}$ according to Eq.\,(\ref{eqn:del_def})
is
\begin{align}
\delta_{1}(\tau=0)=\frac{2\varepsilon^{2}(1-\varepsilon^{2})}{(1+\varepsilon^{2})^{2}}\sum_{n=0}^{\infty}\varepsilon^{2n}(2n(n+1)+1).
\end{align}
Then, assuming the time $\tau$ and spreading parameter $\varepsilon$
are small, we may expand Eqs.\,(\ref{eqn:w1-1})-(\ref{eqn:w2-1})
in terms of these small parameters to find the variance $\delta_{1}$
as a series in orders of $\varepsilon$ and $\tau$. To first order
in $\tau$, $\delta_{1}$ is also first order in $\varepsilon$, 
\begin{align}
\delta_{1}(\tau)=\delta_{1}(\tau=0)+2\varepsilon\tau\left(h_{m}-h_{m-1})\right)+\mathcal{O}(\tau^{2})+\mathcal{O}(\varepsilon^{2}).\label{eqn:del11}
\end{align}
Note that it happens that each order in $\tau$ introduces a factor
of a constant $h_{i}\sim h_{m}$, so we require that $\tau\ll1/h_{m}$
for the expansion to hold. Thus, the growth of the variance $\delta_{1}$
is linearly proportional to the small spreading parameter $\varepsilon$
at short times, and we may provisionally take $\delta_{1}(\tau)=\delta_{1}(0)$,
relying on the smallness of $\varepsilon$. We will check this assumption
numerically below. 

\subsection{Quantum three-wave instability}

We now proceed to solve Eq.\,(\ref{eqn:quant_main}) for a narrowly
distributed initial condition described in Eq.\,(\ref{eq:init_spread}).
Denote by $(\langle n_{1}\rangle,\langle n_{2}\rangle,\langle n_{3}\rangle)=(n_{1i},n_{2i},n_{3i})$,
the initial expected occupation numbers. Letting $\langle n_{1}\rangle=n_{1i}+\delta n_{1}$,
and Taylor expanding Eq.\,(\ref{eqn:quant_main}) around $n_{1i}$,
we have
\begin{align}
\ddot{\delta n_{1}}=2\left(3\delta_{1}(\tau)+n_{2i}n_{3i}-n_{1i}(1+n_{2i}+n_{3i})+\delta n_{1}(2n_{1i}-2n_{2i}-2n_{3i}-1+3\delta n_{1})\right).\label{eq:quant_middle}
\end{align}
Collecting terms allows us to define
\begin{align}
\gamma_{Q} & =2\sqrt{n_{1i}-n_{2i}-n_{3i}-1/2},\label{eqn:omg}
\end{align}
and
\begin{align}
B_{Q}(\tau) & =2n_{1i}(1+n_{2i}+n_{3i})-2n_{2i}n_{3i}-6\delta_{1}(\tau),
\end{align}
similarly to the classical definitions of $\gamma_{C}$ and $B_{C}$,
Eqs.\,(\ref{eqn:class_omg}) and (\ref{eqn:class_gam}). Thus, Eq.\,(\ref{eq:quant_middle})
can be written as
\begin{align}
\ddot{\delta n_{1}}=\delta n_{1}(\gamma_{Q}^{2}+6\delta n_{1})-B_{Q}(\tau).\label{eqn:xtil}
\end{align}
As in the classical case, when $\tau\ll1/\gamma_{Q}$ (or equivalently
$\delta n_{1}\ll\gamma_{Q}^{2}$), the quadratic term may be neglected.
Since the growth of $\delta_{1}$ may be made arbitrarily small with
$\epsilon$, let us assume a constant variance such that $B_{Q}(\tau)=B_{Q}(0)\equiv B_{Q}$.
This results in an approximate quantum three-wave interaction equation,
\begin{align}
\ddot{\delta n_{1}}=\delta n_{1}\gamma_{Q}^{2}-B_{Q},\label{eqn:xtil_2}
\end{align}
which is identical in form to the classical equation, Eq.\,(\ref{eq:class_non_lin}),
and also shares a solution of the same form
\begin{align}
\delta n_{1}=\frac{B_{Q}}{\gamma_{Q}^{2}}+C_{1}e^{\gamma_{Q}\tau}+\left(\frac{B_{Q}}{\gamma_{Q}^{2}}+C_{1}\right)e^{-\gamma_{Q}\tau},\label{eqn:quan_sol}
\end{align}
where $C_{1}$ is a constant determined by the initial conditions.
The above equation for $\delta n_{1}$ implies that the condition
for instability in the quantum theory of the three-wave interaction
(assuming an initial variance which grows slowly in time) is similar
to the classical instability criterion, namely that $\gamma_{Q}^{2}>0$. 

Since it is not possible to formulate a first-order equation for $\langle n_{1}\rangle$
in lieu of the second-order Eqs.\,(\ref{eqn:n1})-(\ref{eq:n3}),
the constant $C_{1}$ must be calculated directly from the Schrödinger
equation at $\tau=0$. Explicitly, 
\begin{align}
\dot{\delta n_{1}}(0)=\partial_{\tau}\langle n_{1}\rangle(0)=\sum_{i=0}^{s_{2}}2\alpha_{i}(0)\dot{\alpha}_{i}(0)(s_{2}-i),
\end{align}
where the time-derivatives of the weights $\dot{\alpha}_{i}$ are
given by Eqs.\,(\ref{eqn:w1-1})-(\ref{eqn:w2-1}). 

\subsection{Classical Correspondence}

In summary, we have found that the quantum theory for three-wave interaction
supports a quantum instability according to Definition \ref{def:1}.
This quantum instability corresponds to the classical description
of the instability, and the unstable solutions are structurally identical.
Both require that the growth rate $\gamma$ is real for an instability,
and both are only applicable so long as the quadratic terms $\delta I_{1}^{2}$
and $\delta n_{1}^{2}$ are small. The growth rates, 
\begin{align*}
\gamma_{Q} & =2\sqrt{n_{1i}-n_{2i}-n_{3i}-1/2},\\
\gamma_{C} & =2\sqrt{I_{1i}-I_{2i}-I_{3i}},
\end{align*}
and other constants of the unstable solutions, 
\begin{align*}
B_{Q} & =2n_{1i}(1+n_{2i}+n_{3i})-2n_{2i}n_{3i}-6\delta_{1}(0),\\
B_{C} & =2I_{1i}(I_{2i}+I_{3i})-2I_{2i}I_{3i},
\end{align*}
differ only by constant factors, and the relative differences between
these constants tend towards zero in the classical limit as $n_{1i},n_{2i},n_{3i}\rightarrow\infty$.
Further, for a fixed spreading parameter $\varepsilon<1$, the initial
condition described in Eq.\,(\ref{eq:init_spread}) will yield a
variance that approaches zero in the classical limit. 

There are crucial differences between the quantum and classical systems
which do not diminish as the photon number increases though. First,
the quantum wave action equation, Eq.\,(\ref{eqn:quant_main}), depends
on the variance of the action, a purely quantum phenomenon. Although
the effect of the variance on the approximate quantum solution may
be reduced if the initial variance is chosen to be small, it still
introduces a new independent parameter which must be chosen carefully
to result in instability. Second, the quantum system does not admit
closed first-order equations for the wave amplitude as the classical
system does in Eq.\,(\ref{eqn:first_order}). This has the effect
of making the constant $C_{1}$ in Eq.\,(\ref{eqn:quan_sol}) non-trivial
to calculate in the quantum system, and since $C_{1}$ depends on
the initial variance, it will not in general converge to the classical
value in the classical limit. Finally, the quantum theory does not
have a zero-energy eigenstate corresponding to the classical equilibrium.
The classical and quantum solutions being compared are the result
of linearizations about an arbitrary initial condition instead of
an equilibrium. 

\subsection{Numerical solution of quantum instability }

In this subsection, we compare the approximate solution of quantum
instability obtained with the numerical solution of the Schrödinger
equation. For a fixed $s_{2}$ and $s_{3}$, we expect the quantum
instability to have the highest growth rate when $(\langle n_{1}\rangle(0),\langle n_{2}\rangle(0),\langle n_{3}\rangle(0)\equiv(n_{1i},n_{2i},n_{3i})=(s_{2},0,0)$
and $s_{2}=s_{3}$. This follows from the definition of the growth
rate of the instability $\gamma_{Q}$ in Eq.\,(\ref{eqn:omg}). Indeed,
in the next section, this case will be used as the example of quantum
instability. For evaluating the validity of the approximate solution
of quantum instability, Eq.\,(\ref{eqn:quan_sol}), this system is
the most favorable because its initial variance $\delta_{1}$ and
spreading parameter $\varepsilon$ are zero. 

\begin{figure}
\includegraphics[width=0.5\linewidth]{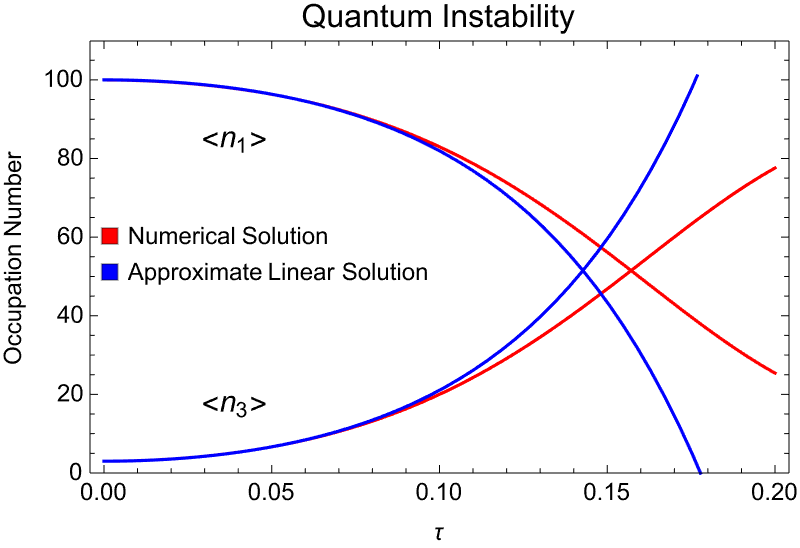} \caption{Numerical solution of the Schrödinger equation, Eq.\,(\ref{eq:schr}),
and approximate linear solution, Eq.\,(\ref{eqn:quan_sol}), of quantum
system with the unstable initial condition $(n_{1i},n_{2i},n_{3i})=(100,10,3)$.
The spreading parameter is $\varepsilon=0.1$, which yields $\delta_{1}(0)=0.02$
and $C_{1}=-3.9$.}
\label{fig:quant_lin}
\end{figure}

Here, we consider a less favorable unstable initial condition $(n_{1i},n_{2i},n_{3i},\varepsilon)=(100,10,3,0.1)$.
Plotted in Fig.\,\ref{fig:quant_lin} are the exact quantum solution
to the Schrödinger equation and the approximate solution of Eq.\,(\ref{eqn:quan_sol})
for that initial condition. Note that while the derivation of the
solution to the linearized Eq.\,(\ref{eqn:xtil_2}) is only valid
until $\tau\sim1/h_{z_{0}}=0.02$ as required by our expansion of
the variance in Eq.\,(\ref{eqn:del11}), the linearized solution
matches the exact solution well beyond that point. At $\tau\sim0.14$,
the condition for the linearization, $\delta n_{1}\ll\gamma_{Q}^{2}$,
breaks as $\delta n_{1}\cong50$ and $\gamma_{Q}^{2}=346$, and the
approximate solution and exact solutions diverge. In the case of $(n_{1i},n_{2i},n_{i0},\varepsilon)=(100,0,0,0)$,
the approximate solution remains valid much longer since the growth
of the variance $\delta_{1}$ is now second order in $\tau$ as shown
in Eq.\,(\ref{eqn:del11}), and the initial condition imposes $\dot{\langle n_{1}\rangle}(0)=0$,
keeping $\delta n_{1}$ smaller than $\gamma_{Q}^{2}$ for much longer. 

\section{Properties of quantum instability\label{sec:Numerical}}

In this section, we investigate the properties of the quantum instability
through numerical solutions. It is demonstrated that the quantum instability
is realized as a cascade of wave functions in the space of occupation
number states. We also show that an instability--admitting Hamiltonian
has a much richer spectrum than a stable Hamiltonian, and quantum
instability is associated with an almost infinite recurrence time.

\subsection{Quantum instability as a wave function cascade in the space of occupation
number states}

\begin{figure}
\includegraphics[width=0.5\linewidth]{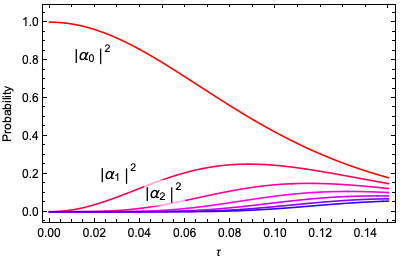}
\caption{Evolution of the probability of occupation number states for unstable
initial condition $(n_{1i},n_{2i},n_{3i})=(100,0,0)$ and spreading
parameter $\varepsilon=0$. This corresponds to $\alpha_{0}=0$ and
$\Psi=\psi_{0}$ at $\tau=0$. The first three states' probabilities
are labeled, and the first seven states are plotted.}
\label{fig:quant_casc}
\end{figure}

The exponential growth in the occupation numbers shown in Eq.\,(\ref{eqn:quan_sol})
and Fig.\,\ref{fig:quant_lin} is realized through a cascade of wave
functions from states with higher $|n_{1}\rangle$ to states with
lower $|n_{1}\rangle$. This cascade is particularly evident with
an initial state which is maximally localized and also maximal in
the expectation value of $\hat{n}_{1}$. Show in Fig.\,\ref{fig:quant_casc}
is such a probability cascade with initial condition $(n_{1i},n_{2i},n_{3i})=(100,0,0)$,
which initializes the $\psi_{0}=|100,0,0\rangle$ state with a probability
1, i.e., $\alpha_{0}=1$ at $\tau=0$. 

\begin{figure}
\begin{minipage}[t]{0.45\columnwidth}%
\includegraphics[width=0.9\linewidth]{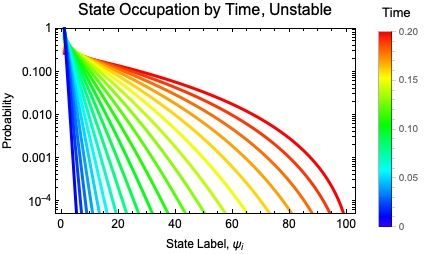}

(a) $(n_{1i},n_{2i},n_{3i})=(100,0,0)$ \label{fig:sfig1}%
\end{minipage}%
\begin{minipage}[t]{0.45\columnwidth}%
\includegraphics[width=0.9\linewidth]{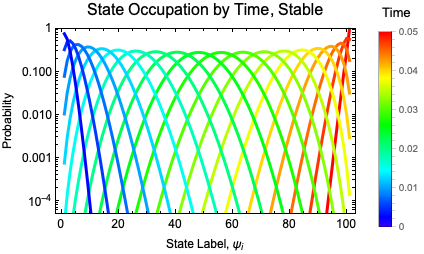}

(b) $(n_{1i},n_{2i},n_{3i})=(100,900,0)$ \label{fig:sfig2}%
\end{minipage}

\caption{Probability distribution over occupation number states at various
times, represented by different colors, for initial condition $\alpha_{0}^{2}(0)=1$.
For the unstable case (a), $s_{2}=s_{3}=100$, while for the stable
case (b), $s_{2}=100$ and $s_{3}=1000$. The unstable initial condition
results in a cascade of probability among all 101 states, and the
variance monotonically increases. For the stable case, the variance
oscillates with a recurrence time of approximately 0.1. Only the first
half of the recurrence time is plotted in (b) for clarity. The unshown
second half would overlap with the first half. }
\label{fig:fig2}
\end{figure}

Only the first seven states' probability evolution is shown in Fig.\,\ref{fig:quant_casc},
but the cascade occurs through all 101 available states as shown in
Fig.\,\ref{fig:fig2}(a). The cascading behavior is characteristic
of the instability in the three-wave interaction. In a stable system,
with an otherwise identical initial probability distribution among
its 101 states, the cascade does not occur as illustrated in Fig.\,\ref{fig:fig2}(b)
for the case of $(n_{1i},n_{2i},n_{3i})=(100,900,0)$. It is interesting
to note that the cascading process evident in Fig.\,\ref{fig:fig2}(a)
is maintained well past the time when the numerical solution and the
approximate linear solution diverge. Also of note in Figs. \ref{fig:fig2}(a)
and \ref{fig:fig2}(b) is the irreversibility of the unstable quantum
system versus the guaranteed reversibility of the stable quantum system.
The recurrence time shown in Fig.\,\ref{fig:fig2}(b) is approximately
$0.1$. 

\subsection{Spectrum property of quantum instability}

\begin{figure}
\includegraphics[scale=0.7]{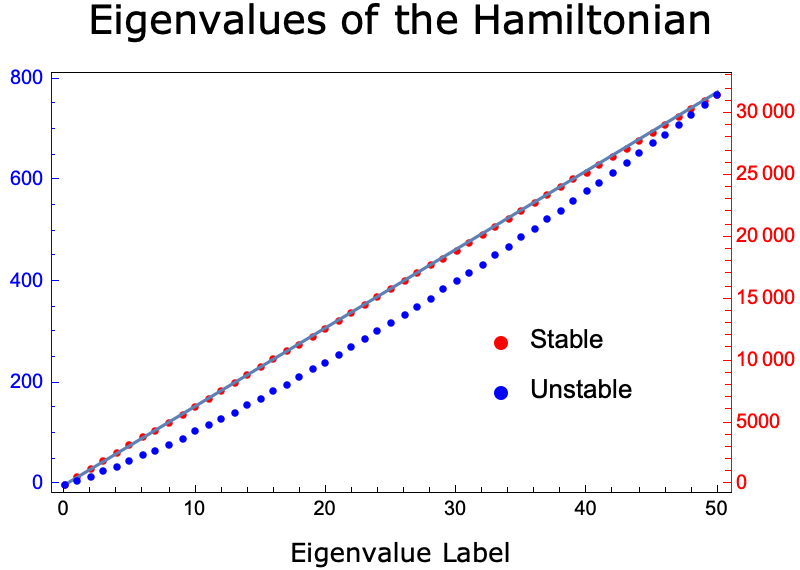}\caption{Eigenfrequencies of quantum systems with (red, upper) and without
(blue, lower) quantum instability. The unstable system has $s_{2}=s_{3}=100$,
while the stable system has $s_{2}=100$ and $s_{3}=1000$. Only 51
of the 101 eigenvalues are shown, since the spectra are symmetric
with respect to the real axis. }

\label{fig:eigen}
\end{figure}

In this subsection, we will look closely at the eigenvectors and eigenvalues
of the two $d=101$ systems shown in Fig.\,\ref{fig:fig2}, with
$(s_{2},s_{3})=(100,100)$ and $(s_{2},s_{3})=(100,1000)$, respectively.
The first system admits quantum instability and the second does not.
According the theoretical analysis developed in Sec.\,\ref{sec:Quantum},
this is because for the second system, $\gamma_{Q}^{2}/4=\langle n_{1}\rangle-\langle n_{2}\rangle-\langle n_{3}\rangle-1/2=3\langle n_{1}\rangle-s_{3}-s_{2}-1/2=3\langle n_{1}\rangle-999.5<0$,
since $\langle n_{1}\rangle=s_{2}-\langle n_{3}\rangle<s_{2}$. The
101 eigenvalues of the two systems are shown in Fig.\,\ref{fig:eigen}.
For the stable system, the eigenvalues are linearly distributed to
a high precision, i.e., 
\begin{equation}
\lambda_{0}=0,\ \lambda_{1}\approx630,\ \lambda_{2}=-\lambda_{1},\ \lambda_{3}\approx2\lambda_{1},\dots\lambda_{100}=-50\lambda_{1},
\end{equation}
while it can be seen in Fig.\,\ref{fig:eigen} that the eigenvalues
in the unstable case are nonlinearly distributed. For intermediate
values of $s_{3}>s_{2}$ and $s_{3}<1000$, it was found that the
eigenvalues lie between these two extremes. 

The linearity of the eigenvalues of for the stable case places strong
limitations on the allowable dynamics of the system. To see this directly,
we analyze the frequency decomposition of the expectation of the occupation
numbers. Denote by $\mathbf{v}_{j}$ the eigenvectors of $H.$ In
the $\psi_{j}$ bases, 
\begin{equation}
\mathbf{v}_{j}=\sum_{k=0}^{s_{2}}\beta_{jk}\psi_{k},
\end{equation}
which defines the transformation matrix $\beta_{jk}$. For a state
$\Psi(t)$, we have
\begin{equation}
\Psi(t)=\sum_{j=0}^{s_{2}}\alpha_{j}(t)\psi_{j}=\sum_{j=0}^{s_{2}}\epsilon_{j}\mathbf{v}_{j}e^{-i\lambda_{j}t}.
\end{equation}
This then identifies
\begin{equation}
\alpha_{j}(t)=\sum_{k=0}^{s_{2}}\epsilon_{k}\beta_{kj}e^{-i\lambda_{k}t},
\end{equation}
The expectation $\langle n_{3}\rangle(t)$ can be evaluated as
\begin{align}
\langle n_{3}\rangle & =\sum_{j=0}^{s_{2}}|\alpha_{j}(t)|^{2}j\label{eqn:n1_def}\\
 & =\sum_{j=0}^{s_{2}}\left|\sum_{k=0}^{s_{2}}\epsilon_{k}\beta_{kj}e^{-i\lambda_{k}t}\right|^{2}j\\
 & =\sum_{j=0}^{s_{2}}\sum_{k=0}^{s_{2}}|\epsilon_{k}\beta_{kj}|^{2}j\nonumber \\
 & \ \ \ \ \ +\sum_{j=0}^{s_{2}}\left\{ \sum_{k=1}^{s_{2}}\left[\epsilon_{0}\epsilon_{k}\beta_{0j}^{\dagger}\beta_{kj}e^{i(\lambda_{0}-\lambda_{k})t}+c.c.\right]\right.\nonumber \\
 & \ \ \ \ \ \ \ \ \ \ +\sum_{k=2}^{s_{2}}\left[\epsilon_{1}\epsilon_{k}\beta_{1j}^{\dagger}\beta_{kj}e^{i(\lambda_{1}-\lambda_{k})t}+c.c.\right]+\nonumber \\
 & \ \ \ \ \ \ \ \ \ \ +\dots\nonumber \\
 & \ \ \ \ \ \ \ \ \ \ +\epsilon_{s_{2}-1}\epsilon_{s_{2}}\beta_{s_{2}-1,j}^{\dagger}\beta_{s_{2}j}e^{i(\lambda_{s_{2}-1}-\lambda_{s_{2}})t}+c.c.\biggr\}.\label{eq:n3_expansion}
\end{align}
The spectral frequencies available for $\langle n_{3}\rangle(t)$
are the differences between each pair of the eigenfrequencies $\lambda_{i}-\lambda_{j}$
where $i\neq j$. The weights are determined by the weighting of eigenvectors,
$\epsilon_{i}$, and the transformation matrix $\beta_{kj}$. 

For the stable quantum Hamiltonian with $(s_{2},s_{3})=(100,1000)$
defined above, the linear spacing of its eigenvalues means that there
are only 101 spectral frequencies (corresponding to combinations of
the 50 distinct eigenvalue absolute values and the 0 eigenvalue) available
for $\langle n_{3}\rangle(t)$. It also implies that its spectrum
constitutes a Fourier series. By contrast, the unstable Hamiltonian
with $(s_{2},s_{3})=(100,100)$ has $\text{Floor}(d^{2}/4)+1=2551$
spectral frequencies available. The values of the spectral frequencies,
not just their quantity, are also important. The maximum recurrence
time, when the system will begin to repeat itself, in either system
will be the least common multiple of the spectral periods. For the
stable Hamiltonian, this value is guaranteed to exist, since the Fourier
periods will all be rational multiples of a single, base period. Conversely,
we expect an exact recurrence time to never exist in the unstable
system. Exact analytical solutions to the eigenvalues of even low-dimensional
Hamiltonians show that the spectral frequencies are incompatibly irrational,
implying an almost infinite recurrence time. This irreversibility
is a familiar hallmark of instability in a classical system. 

\begin{figure}
\noindent\begin{minipage}[t]{1\columnwidth}%
\subfloat[Amplitude of eigenvectors, $\epsilon_{i}$, for the stable system.]{\includegraphics[width=0.45\linewidth]{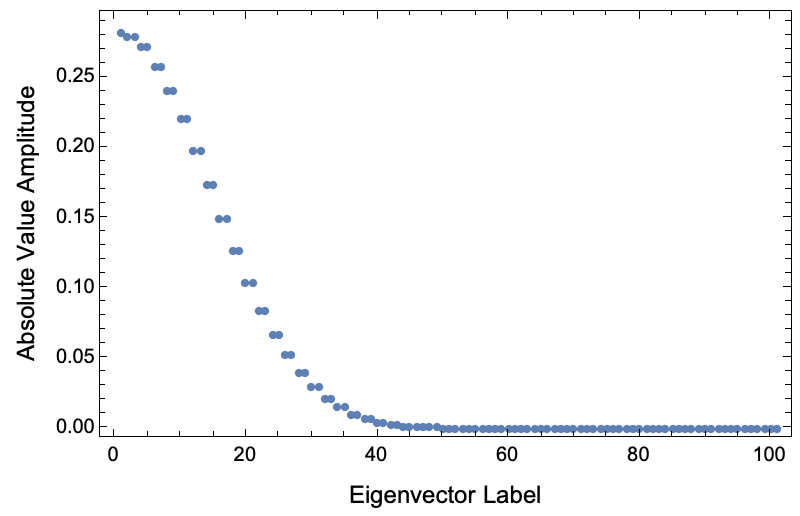}

}\hfill{}\subfloat[Spectrum weight of $\langle n_{3}\rangle(t)$ for the stable system.]{\includegraphics[width=0.45\linewidth]{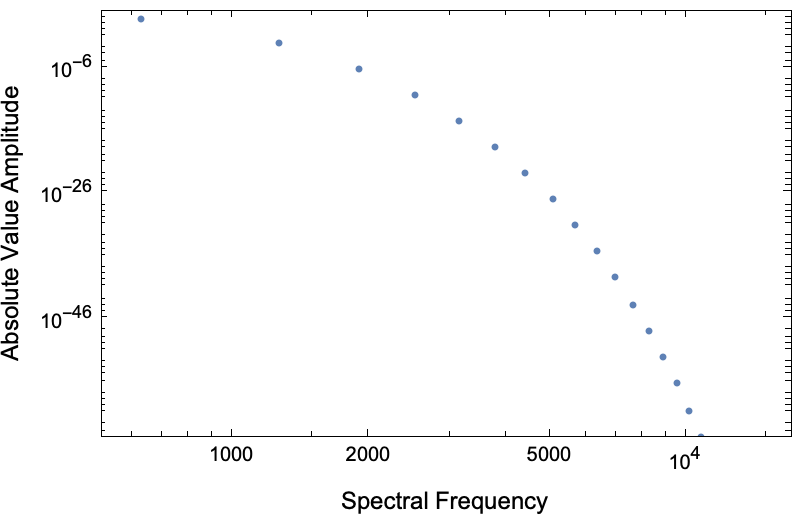}}%
\end{minipage}\smallskip{}
\noindent\begin{minipage}[t]{1\columnwidth}%
\subfloat[Amplitude of eigenvectors, $\epsilon_{i}$, of the unstable system.]{\includegraphics[width=0.45\linewidth]{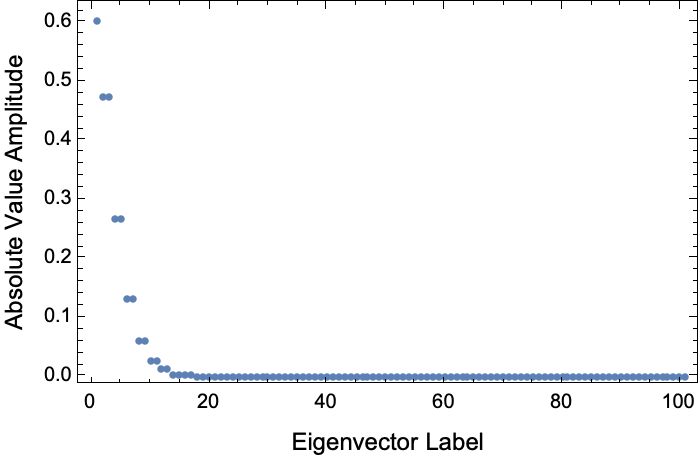}}\hfill{}\subfloat[Spectrum weight of $\langle n_{3}\rangle(t)$ for the unstable system. ]{\includegraphics[width=0.45\linewidth]{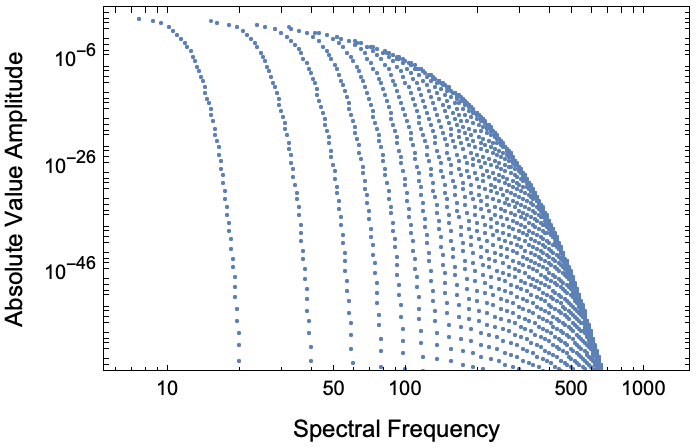}}%
\end{minipage}\caption{Eigenvector weights $\epsilon_{i}$ and spectrum weight of of $\langle n_{3}\rangle(t)$
(the coefficients $\epsilon_{i}\epsilon_{j}\beta_{ik}\beta_{jk}$
for each frequency $\lambda_{i}-\lambda_{j}$ in Eq.\,(\ref{eq:n3_expansion}))
for the stable $(s_{2},s_{3})=(100,1000)$ and unstable $(s_{2},s_{3})=(100,100)$
systems. The initial condition is $\alpha_{0}=1$, which corresponds
to the $\psi_{0}=|100,0,0\rangle$ mode having probability 1. Most
of the 101 and 2551 spectral modes have zero amplitude in (b) and
(d), respectively, and are not displayed. }

\label{fig:quant_unst_stb}
\end{figure}

The amplitude of eigenvalues and frequency spectrum of the stable
and unstable systems given the initial condition $\alpha_{0}(0)=1$
are plotted in Fig.\,\ref{fig:quant_unst_stb}. The unstable system
is $(s_{2},s_{3})=(100,100)$ with $(n_{1i},n_{2i},n_{3i})=(100,0,0)$,
and the stable system is $(s_{2},s_{3})=(100,1000)$ with $(n_{1i},n_{2i},n_{3i})=(100,900,0)$.
The spectral differences between the stable and unstable quantum systems
discussed above are evident. 

\section{Discussion and Conclusions\label{sec:Conclusions}}

For a static, finite-dimensional Hermitian quantum system, all eigenfrequencies
of the system are real, and the dynamics is unitary. However, this
does not preclude exponential growth of the expectations of observables
that do not commute with the Hamiltonian. In the present study, quantum
instability is defined as a solution for which the expectation of
an observable deviates exponentially relative to an initial value.
The quantum theory of the three-wave interaction, obtained by quantizing
the classical Hamiltonian using a field-theoretical method, maps the
nonlinear classical Hamiltonian into a finite-dimensional Hermitian
system. We have shown that the quantum theory admits a quantum instability
corresponding to the classical three-wave instability. The quantum
and classical descriptions of the three-wave instability require the
same conditions to occur, with the quantum theory having additional
requirements on the variance of its initial condition. In the classical
limit, both theories predict the same growth rate. We numerically
demonstrated that this quantum instability is realized as a cascade
of wave functions in the space of occupation number states, and further
showed that such a cascade does not occur with a stable Hamiltonian.
It is possible that other instabilities described by classical theory,
especially the opto-mechanical instability \citep{Ludwig_2008,bennett,rodriguez},
are also realized quantum mechanically through a cascade of wave functions
in the space of occupation number states. The Hamiltonian of an unstable
quantum system is shown to possess a much richer spectrum than the
Hamiltonian of a stable quantum system. Additionally, the unstable
quantum system exhibits irreversibility, while the stable system has
a relatively small recurrence time. 

Future work would aim to characterize other quantum instabilities
and develop a general framework for their recognition and correspondence
with classical systems. It is possible that the three-wave interaction
and other physical processes may admit quantum instabilities which
have no classically unstable counterparts. Finding these native quantum
instabilities would provide stronger motivation for the definition
of quantum instability proposed in the present study. 

Current technology allows for the quantum three-wave instability to
be simulated using a quantum hardware. The work performed by Shi \textit{et
al.} in simulating the quantum three-wave interaction utilized only
two qubits and three of their four possible states, $|00\rangle$,
$|01\rangle$, and $|10\rangle$, to represent $d=s_{2}+1=3$ states
\citep{shiPhysRevA}. Although the linear solution given by Eq.\,(\ref{eqn:quan_sol})
could have been simulated, the growth rate of the quantum instability
would have been too small to notice at such a low dimension. Also,
the dimension was too small to compare with the classical instability.
However, since the number of states representable by $n$ qubits is
$d\propto2^{n}$, quantum hardware with sufficient numbers of qubits
to simulate the quantum instability and compare to the classical instability
already exist \citep{rigetti}. Of course, this issue is complicated
by the unitary operator needing $d^{2}=2^{n^{2}}$ gates to be approximated
using the gates available to the system. Though the method of implementing
a single, special-made gate as in \citep{shiPhysRevA} somewhat mitigates
this problem.
\begin{acknowledgments}
This research was supported by the U.S. Department of Energy (DE-AC02-09CH11466).
\end{acknowledgments}

\bibliographystyle{apsrev4-2}
\bibliography{draft}

\end{document}